\documentclass[conference]{IEEEtran}
\IEEEoverridecommandlockouts
\usepackage{cite}
\usepackage{amsmath,amssymb,amsfonts}
\usepackage{algorithmic}
\usepackage{float}
\usepackage{graphicx}
\usepackage{xcolor}
\usepackage{tabularx}
\usepackage{multirow}
\usepackage{lipsum}
\usepackage{enumitem}
\usepackage{booktabs}
\usepackage{ragged2e}

\usepackage{gensymb}
\usepackage{amsmath}

\graphicspath{ {figures/} }
\usepackage{filecontents}
\newcolumntype{Y}{>{\RaggedRight\arraybackslash}X} 
\def\BibTeX{{\rm B\kern-.05em{\sc i\kern-.025em b}\kern-.08em
    T\kern-.1667em\lower.7ex\hbox{E}\kern-.125emX}}
\begin{document}

\title{Performance Analysis of LOS THz Systems under Misalignment and Deterministic Fading}

\author{
 \IEEEauthorblockN{1\textsuperscript{th} Rayyan Abdalla}
\IEEEauthorblockA{\textit{Department of Electrical and Computer Engineering} \\
\textit{Arizona State University}\\
rsabdall@asu.edu}
\and
\IEEEauthorblockN{2\textsuperscript{th} A.Brinton Cooper III}
\IEEEauthorblockA{\textit{Department of Electrical and Computer Engineering} \\
\textit{Johns Hopkins University}\\
abcooper@jhu.edu}
}
\maketitle

\begin{abstract}

Line-of-sight (LOS) wireless communication at terahertz (THz) frequency bands is envisioned to play a major role in defining next-generation wireless technologies. This work analyzes the performance of a potential LOS THz system experiencing propagation loss and misaligned antenna beams. The THz channel particularities are discussed in terms of deterministic path loss, molecular absorption effect and stochastic fading due to antenna pointing errors. Assuming phase shift keying (PSK) modulation schemes, simplified analytical expressions are approximated for computing symbol error rate (SER) of the proposed THz system. Monte Carlo simulations are applied to verify theoretical model accuracy over various transmission distances and misalignment scenarios. The derived SER formulas match simulation results for Signal-to-noise ratio (SNR) above 35 dB at transmission distance up to 100 m and antenna displacement jitter variance of 0.05 $m^2$. In general, the theoretical model mismatch does not exceed 2 dB for lower SNR levels.

\end{abstract}


\section{Introduction}\label{introduction}
Future wireless technologies are anticipated to meet the demand for bandwidth consuming services and achieve ultra-fast low-latency wireless transmission. Emerging indoor and outdoor wireless applications require short- to medium-range data communications at a rate scale up to Terabit per second (Tbps). Therefore, the telecom world has recently considered utilizing millimeter-wave (mmWave) bands for fifth-generation (5G) systems and beyond. Nonetheless, the foreseen quality of service and user experience requirements are beyond the capabilities of currently deployed mm-Wave systems. These limitations motivated investigating higher frequency bands, particularly in the THz spectrum (0.1-10 THz) in order to secure unmatched bandwidth increase and achieve ultra-high transmission rates \cite{chen2019survey,liu2021thz}.  THz devices are seen to establish effective alternatives for wireless fiber extenders, multi-hop wireless backhaul links, indoor local-area-networks (LANs) and several other applications. \par
  Communications over THz bands encounter several challenges due to high frequency range and distinct channel characteristics as opposed to lower-frequency channels. THz wave propagation suffers severe attenuation due to deterministic free-space path loss and molecular absorption caused by water vapor and variant atmospheric conditions. This has destructive effect on the transmitted electromagnetic (EM) signals as their energy is transformed to medium molecules \cite{kokkoniemi2016discussion}. That is, such channel attenuation imposes the employment of high-gain antennas with narrow directive beams to ensure reliable energy transmission between THz transmitter and receiver. However, antenna beams may not be fully aligned due to fluctuations induced by environmental turbulence, which cause antenna swaying and thus pointing errors. The resultant pointing errors severely degrade link quality and may result in system outage. Though other factors, such as hardware impairments and small-scale fading may affect THz link performance \cite{papasotiriou2020performance}, path propagation loss and the random antenna beam misalignment are major fading contributors to LOS THz systems.\par 
 Several deterministic and stochastic models addressing the THz channel have been proposed in the literature. The free space path loss (FSPL), particularly diffraction and absorption loss, was studied by \cite{kokkoniemi2016discussion,oyeleke2020absorption}. A simplified approach modeling molecular absorption as a function of frequency and atmospheric conditions was provided by \cite{kokkoniemi2018simplified}. The authors in \cite{taherkhani2020performance} quantified pointing error effect using conventional channel models of Gamma-Gamma, Lognormal and Weibull distributions. In \cite{dabiri2022pointing}, the authors modeled misalignment as a random factor affecting antenna gains. Another work considers more comprehensive misalignment model \cite{farid2007outage}, originally derived for free space optical (FSO) links, and which accounts for atmospheric turbulence, receiver detector size, displacement jitter, this model was also adopted by \cite{boulogeorgos2019analytical,kokkoniemi2020impact}. In all aforementioned models, THz systems performance is assessed in terms of outage probability and link capacity. To the best of authors knowledge, performance of digital modulation schemes over LOS THz channel has not been covered in the open literature.\par 
 In this work, a THz communication system model is approximated, which includes the joint effect of misalignment and path propagation loss. The approximated system assumes minimal phase noise and hardware imperfections. Assuming PSK modulation, novel closed-form analytical expressions are derived for both instantaneous and average symbol error rates. Accuracy is attested by carrying out Monte Carlo simulations for different link parameter settings. The organization is as follows: section \ref{sec:system model} presents system and channel model. Analytical performance evaluation is covered in section \ref{sec:performance}. Section \ref{results} discusses simulation results and concluding remarks are given in section \ref{con}.

\section{System Model}
\label{sec:system model} 
 This study considers a LOS point-to-point (P2P) THz link model with highly-directive, yet turbulent, transmit (TX) and receive (RX) antennas. The TX antenna emits a Gaussian beam onto a lens installed at the RX feed detector. Transmission is assumed over a flat fading subcarrier propagating through a channel \textit{h} with additive white Gaussian noise (AWGN) $ n \sim CN (0,\sigma_n^2) $. For  information signal \textit{s} and received signal \textit{y}, the equivalent baseband model equation is
 \begin{equation} \label{system_equ} 
     y = hs + n
 \end{equation}

All system components are statistically independent. The complex AWGN has a zero-mean and a variance \textit{$\sigma_n^2$}.The channel gain results due to three factors, which are: FSPL \textit{$h_p$}, molecular absorption loss \textit{$h_a$} and misalignment fading effect \textit{$h_m$}. 
\begin{equation} \label{channel} 
     h=h_p h_a h_m
\end{equation}
It should be noted that both FSPL and molecular absorption gains comprises channel deterministic attenuation, while power loss due to pointing errors and misalignment is modeled randomly. The following subsections substantially analyze channel components and presents general rules for implementing THz channel model.

\subsection{FSPL Attenuation}
Common LOS path loss is unavoidable for EM signal propagation, according to Friis transmission equation \cite{balanis2011modern}
\begin{equation} \label{path_loss} 
     h_p = \frac{c \sqrt{G_{TX} G_{RX}}}{4 \pi df}
\end{equation}
Where \textit{$G_{RX}$} and \textit{$G_{TX}$} are RX and TX antenna gains, \textit{$c$} is the speed of light, \textit{$f$} is carrier frequency and \textit{$d$} is the distance between the TX and RX antennas. It is evident that the FSPL is much smaller for THz frequencies than mmWave and Microwave frequencies. This manifest the necessity of high antenna gains to mitigate THz spreading loss.

\subsection{Molecular Absorption Loss}
Energy dissipation due to molecular absorption is one of the major fading sources that affects the THz channel. Atmospheric molecules, such as water vapour and oxygen, partially absorb the energy of emitted THz signals. Consequently, this causes resonant peaks and splits THz frequency into multiple transmission windows \cite{liu2021thz,oyeleke2020absorption}. As defined by Beer-Lambert law, the absorption loss grows exponentially with distance according to 
\begin{equation}\label{abs_equ}
    h_a = e^{-\frac{1}{2}k_{a}(f)d}
\end{equation}
Where \textit{$k_a(f)$} is the absorption coefficient, which indicates the area per unit volume within which medium molecules absorb EM signal energy. Determination of the absorption coefficient for high frequencies is a little involved and requires consideration of atmospheric pressure, relative humidity and medium temperature. Nonetheless, a simplified absorption model, derived by  \cite{kokkoniemi2018simplified}, gives an accurate coefficient estimation for a transmission window between 200 to 400 GHz frequency band and up to one kilometer link distance. Following the results of the corresponding model

\begin{equation}\label{egu1}
k_{a}(f)= g(f)+y_1(f,\nu) +y_2(f,\nu)
\end{equation}
and the terms \textit{$y_1(f,\nu)$}, \textit{$y_2(f,\nu)$} and \textit{$g(f)$} are evaluated as
\begin{equation}\label{g}
g(f)= c_0+c_1f+c_2 f^2+c_3f^3
\end{equation}

\begin{equation}\label{y1}
y_1(f,\nu)=\frac{0.2205\nu(0.1303\nu+0.0294)}{{(0.4093\nu+0.0925)}^2+{\left(\frac{100f}{c}-10.835\right)}^2} 
\end{equation}

\begin{equation}\label{y2}
y_2(f,\nu)=\frac{2.014\nu(0.1702\nu+0.0303)}{{(0.537\nu+0.0956)}^2+{\left(\frac{100f}{c}-12.664\right)}^2} 
\end{equation}

Where $c_0 =(-6.36\times10^{-3}) $, $c_1 = (9.06\times10^{-14} Hz^{-1})$, $c_2 = (-3.94\times10^{-25} Hz^{-2})$, $c_3 = (5.54\times10^{-37} Hz^{-3})$ and $\nu$ is the volume mixing ratio of the water vapour which is given by:
\begin{equation}\label{nu}
    \nu = \frac{\phi}{100}\frac{p_w(T,p)}{p}
\end{equation}

Where \textit{$\phi$} is the relative humidity, $p$ is the pressure and \textit{$p_w(T,p)$} is the saturated water vapour partial pressure, which can be calculated as \cite{alduchov1996improved}
\begin{equation}\label{pw}
   p_w(T,p)= q_1(q_2+q_3p_{h})\exp{\left( \frac{q_4(T-q_5)}{T-q_6} \right)}
\end{equation}
Where $q_1 = 6.1121$, $q_2 = 1.0007$, $q_3 =3.46\times10^{-6} hPa^{-1} $, $q_4 = 17.502$ , $q_5=273.15$ \degree$K $ and $q_6 = 32.18 $\degree$K$. Note that \textit{$p_h$} is the pressure in hectopascal and \textit{T} is the absolute temperature. This model demonstrates relatively accurate representation of absorption loss under moderate conditions of 
 1 atm pressure, temperature 296\textit{$\degree$K} and relative humidity of 50\% \cite{boulogeorgos2019analytical} 

\subsection{Misalignment Fading}
\begin{figure}[!htb]
    \centering
    \includegraphics[height=5.5cm,width=7cm]{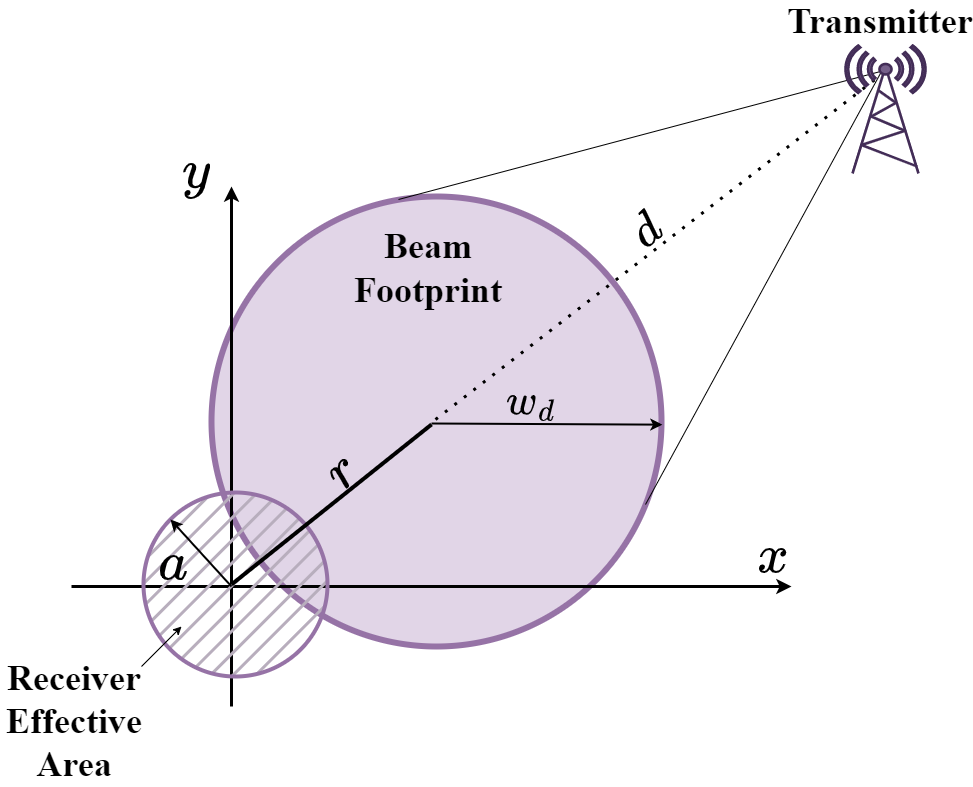}
    \caption{Transmitter beam footprint and receiver detector plane with misalignment.}
    \label{fig:misalign}
\end{figure}

Directive antennas of THz transceivers are considerably susceptible to environmental effects and buildings sways, which results in pointing errors and misaligned antenna beams. A realistic statistical model proposed by \cite{farid2007outage} is utilized to address pointing error distribution and quantify the impact of beam misalignment. As depicted in Fig. \ref{fig:misalign}, assume a receiver has circular aperture area of radius \textit{a}. The transmitter radiates a symmetric beam that has a waist \textit{$w_d$} at distance \textit{d}. The pointing error \textit{r} is expressed as the radial distance between reception area and transmitter beam focus centers. \par

\begin{figure}[!htb]
    \centering
    \includegraphics[scale=0.18]{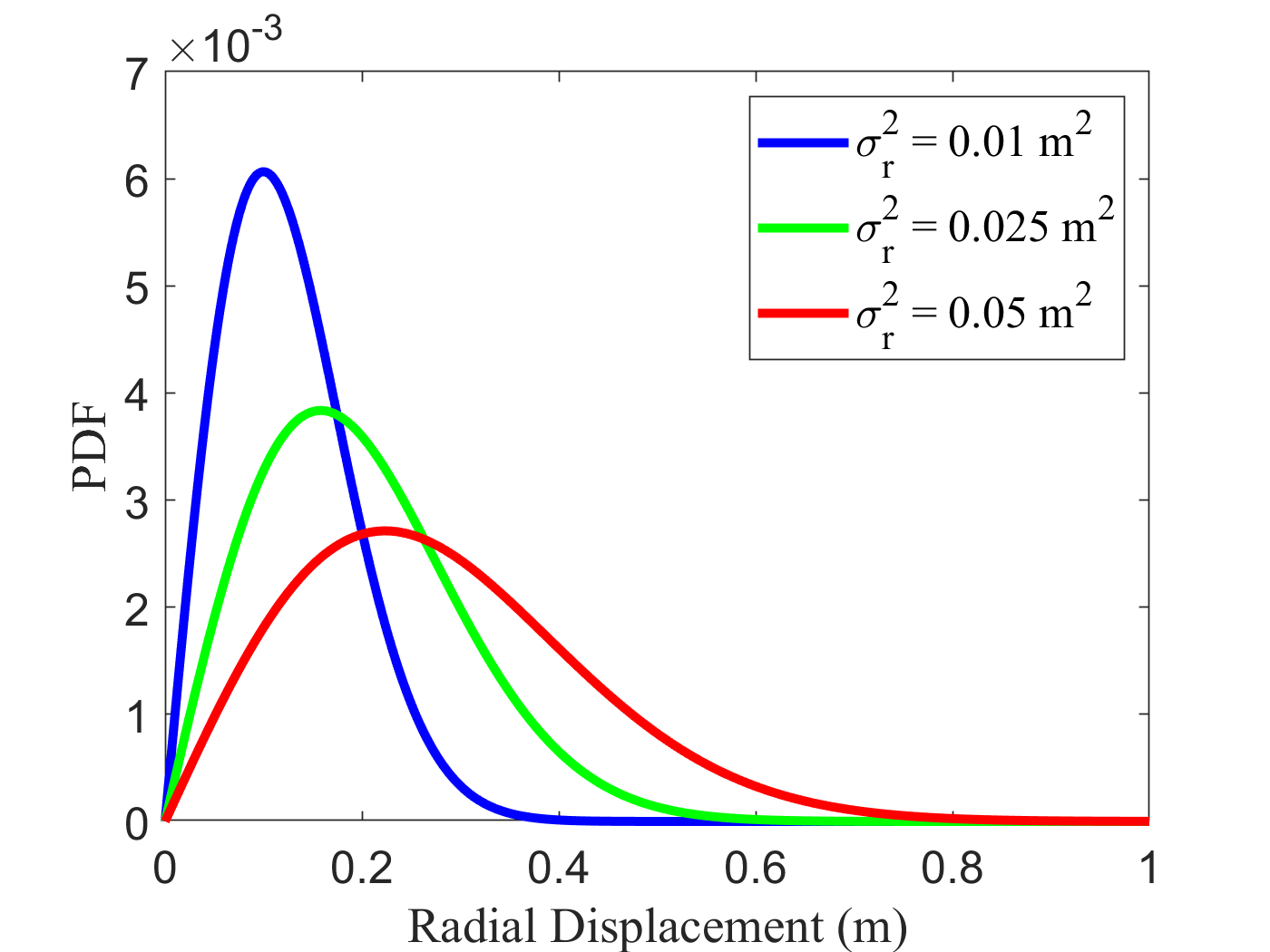}
    \caption{PDFs of pointing error misalignment for jitter variances 0.01, 0.025 and 0.05 $m^2$}
    \label{fig:pdfs}
\end{figure}

It is noted that one-dimensional (1D) antenna displacement follows Gaussian distribution \cite{kokkoniemi2020impact}. Therefore, radial pointing error \textit{r} is resultant due to two 1D displacement components and hence Rayleigh distributed with jitter variance \textit{$\sigma_r$}. Figure \ref{fig:pdfs} shows the probability density function (PDF) of radial displacement for different jitter variances.\par
According to \cite{farid2007outage}, the equivalent beam width that takes into account the receiver radius is given by

\begin{equation}\label{weq}
    w_{eq}^2 = w_d^{2} \frac{\sqrt{\pi} \mbox{erf}(u)}{2u\exp{(-u^2)}},   u=\frac{\sqrt{\pi}a}{\sqrt{2}w_d}
\end{equation}
The misalignment gain in the presence of pointing error can then be approximated by
\begin{equation}\label{hm}
    h_m(r) \approx A_0 \exp{\left (-\frac{2r^{2}}{w_{eq}^2}\right)}
\end{equation}

Where $A_0={ \left(\mbox{erf}(u)\right)}^2$ is fraction of power collected by the receiver when $r=0$. Since $r$ is  a Rayleigh random variable, the PDF of misalignment fading can be written as \cite{farid2007outage,boulogeorgos2019analytical}
\begin{equation}\label{fhm}
    f_{h_m}(x) = \frac{\gamma^2}{A_0^{\gamma^2}}x^{\gamma^{2}-1},  0 \leq x\leq A_0
\end{equation}
The parameter $\gamma = w_{eq}/2\sigma_r$ represents the ratio between equivalent beam width and the standard deviation of pointing error at the receiver. It is interesting to note that this misalignment model is applicable to wide range of frequencies where narrow symmetric antenna beams are used for transmissions \cite{papasotiriou2020performance,farid2007outage,kokkoniemi2020impact}.

\section{Performance Analysis}\label{sec:performance}

\label{sec:section3}
This section presents a simplified theoretical framework for performance assessment of the proposed system model. The instantaneous SNR expression is obtained to derive novel closed-form expressions for average SER of uncoded single-input-single-output (SISO) THz system. Since THz transceivers are normally restricted to modulation schemes with lower spectral efficiency \cite{miretti2022little}, binary phase shift keying (BPSK) and quadratic phase shift keying approaches are assumed for symbol modulation.\par

 Referring to (\ref{system_equ}), for a transmitted symbol power $E_s$, the instantaneous SNR, of the THz system is given by
\begin{equation}\label{snri}
   \mbox{SNR}_{inst} = |h|^{2}\overline{\rho}, \overline{\rho}=\frac{E_s}{\sigma_n^2}
\end{equation}

Where $\overline{\rho}$ is the average SNR of an uncoded communication system under AWGN with negligible quantization error.

\subsection{BPSK Performance}
In a BPSK system with transmitted signal $s \in \{\sqrt{E_s}, - \sqrt{E_s}\}$, the instantaneous symbol error probability of a BPSK-modulated system, $P_{e_{BPSK}}$ is evaluated as \cite{goldsmith2005wireless, meghdadi2008ber}

\begin{equation}\label{pe}
   P_{e_{BPSK}} = Q\left( \sqrt{2 \times \mbox{SNR}_{inst}}\right)
\end{equation}
Where $Q(.)$ is the tail distribution function of standard Gaussian distribution. By considering the joint effect of the THz channel and AWGN on a BPSK-modulated signal, The instantaneous SER of the BPSK THz system is
\begin{equation}\label{pe1}
   P_{e_{BPSK}} = Q\left( \sqrt{2 |h_p h_a h_m|^{2}\overline{\rho} }\right)
\end{equation}

 It can be seen that the instantaneous system performance is a random function of the THz channel fading gains. The average symbol error rate (SER) can be found by obtaining the mean value of the instantaneous system error. Since $h_m$ is the only random fading component, average SER equation is
\begin{equation}\label{pe_dash}
   \overline{P}_{e_{BPSK}} = E_{f_{h_m}} \left[Q\left( \sqrt{2|h_p h_a|^{2}| h_m|^{2}\overline{\rho}}\right) \right]
\end{equation}
Where $E_{f_{h_m}}[.]$ is the expected value with respect to misalignment PDF given in (\ref{fhm}). That is
\begin{equation}\label{pe_2}
   \overline{P}_{e_{BPSK}} =  \frac{\gamma^2}{A_0^{\gamma^2}} \int_{0}^{A_0} Q\left( \sqrt{2\overline{\rho}|h_p h_a|^{2} x^{2}}\right) x^{\gamma^{2}-1} \,dx 
\end{equation}
A closed-form evaluation of the integral (\ref{pe_2}) requires suitable approximation of  the Gaussian $Q$ function. While several tight bounds are available for approximating the normal tail function, one should consider applicability range and convergence interval. Consequently, this work utilizes Chernoff exponential bound discussed in \cite{1210748}, which is 
\begin{equation}\label{pe_22}
  Q(x) \leq \frac{1}{2} \exp{\left(- \frac{1}{2}x^{2}\right)},  x > 0
\end{equation}
Substituting (\ref{pe_22}) into (\ref{pe_2}) and recalling (\ref{fhm}), a tractable formula to compute average SER is obtained by
\begin{equation}\label{ser}
   \overline{P}_{e_{BPSK}} = \frac{\mathcal{B}}{2 \mathcal{A}^{\mathcal{B}}} \Gamma\left( \mathcal{A}, \mathcal{B}\right)
\end{equation}

Where
\begin{equation} \label{AB}
    \mathcal{A} = \overline{\rho} A_0^{2}{|h_p h_a|}^2 , \hspace{0.3in} \mathcal{B} = \frac{\gamma^2}{2}
\end{equation}

 and the function
 $\Gamma(.,.)$ is the lower incomplete Gamma function.

\subsection{QPSK Performance}
The QPSK modulation scheme applies two BPSK modulation on in-phase and quadrature
components of the signal. The instantaneous symbol error rate of the QPSK  system, $P_{e_{QPSK}}$,  is the probability of either branch has an error \cite{goldsmith2005wireless, meghdadi2008ber} and is given by
\begin{equation}\label{peqpsk}
   P_{e_{QPSK}} = 1-  \left[1-Q\left( \sqrt{\mbox{SNR}_{inst}}\right) \right]^2
\end{equation}

Therefore, substituting the expression for $\mbox{SNR}_{inst}$ and obtaining average SER of the QPSK THz system as:
\begin{equation}\label{peq}
  \overline{P}_{e_{QPSK}}  = 
  E_{f_{h_m}} \left[ 1-  \left[1-Q\left( \sqrt{|h_p h_a h_m|^{2}\overline{\rho}}\right) \right]^2 \right]
\end{equation}

Following the same approach of (\ref{pe_dash}) and making use of (\ref{pe_22}), an exact formula of average SER of the THz QPSK system is found to be

\begin{equation}\label{ser_qpsk}
   \overline{P}_{e_{QPSK}} = \frac{ 2^{\mathcal{B}} \mathcal{B}}{ \mathcal{A}^{\mathcal{B}}} \Gamma\left( \frac{\mathcal{A}}{2}  , \mathcal{B}\right) - \frac{\mathcal{B}}{4 \mathcal{A}^{\mathcal{B}}} \Gamma\left( \mathcal{A}, \mathcal{B}\right)
\end{equation} 
Where $\mathcal{A}$ and $\mathcal{B}$ are given in (\ref{AB}).\par
 With a prior knowledge of system parameters, the equations (\ref{ser}) and (\ref{ser_qpsk}) give relatively accurate estimation of BPSK and QPSK THz link performance, respectively. This is to be demonstrated with simulation results for applicable ranges of fading coefficients.

\section{Simulation and Results Discussion}\label{results}
In this section, the THz system performance is investigated by means of analytical and Monte Carlo simulation results. The accuracy of the derived average SER formulas is demonstrated for variable system parameters to determine applicability range. System parameters for computer simulations are provided in Table \ref{tab:table}. Unless otherwise specified, the default transmission distance is 50 m and pointing error variance is 0.01 $m^2$.

\begin{table}[!htb]
 \caption{Basic system parameters for simulation}
  \centering
  \begin{tabular}{l|l|l}
    \cmidrule(r){1-3}
    \textbf{Parameter}     & \textbf{Description}     & \textbf{Value} \\
    \midrule
    $f$ & THz carrier frequency  & 300 GHz     \\
    $G_{TX}$     & TX antenna gain & 55 dBi      \\
    $G_{RX}$     & RX antenna gain       & 55 dBi  \\
      $d$     & TX-RX distance       & 30 - 80 m  \\
       $T$     & Medium temperature      &  296 \textit{$\degree$K}\\
       $p$     & Atmospheric pressure       & 101325 Pa  \\
       $\phi$     & Relative humidity       & 50\%  \\
       $a$     & RX detection area radius       & 10 cm  \\
       $w_d$     & TX beam footprint radius       & 60 cm  \\
       $\sigma_r$     &  Jitter variance      & 0.01 - 0.05 $m^2$  \\
    \bottomrule
  \end{tabular}
  \label{tab:table}
\end{table}

\begin{figure}[!htb]
    \centering
    \includegraphics[scale=0.2]{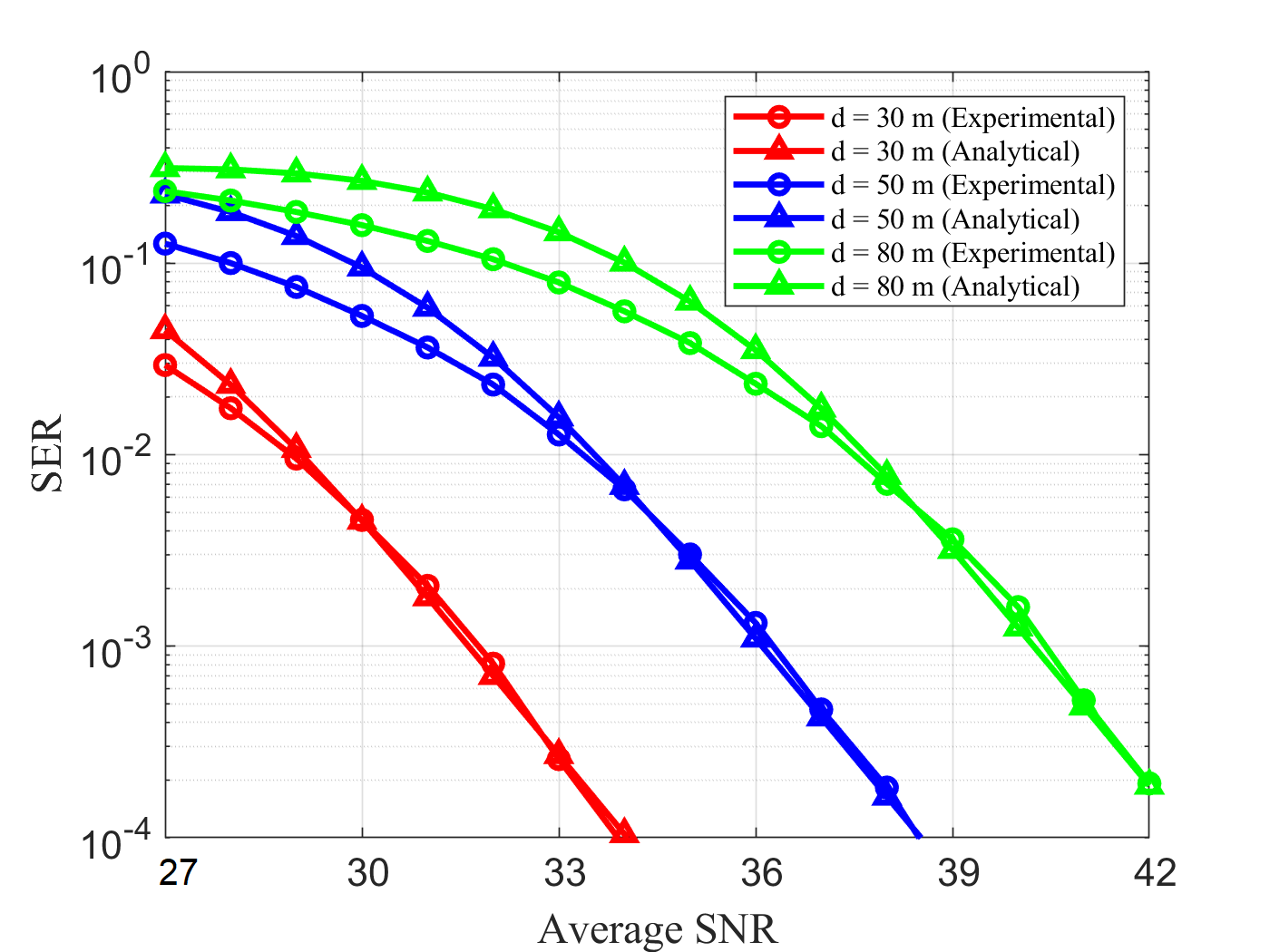}
    \caption{BPSK average SER vs. Average SNR assuming jitter variance of 0.01 $m^2$}
    \label{fig:dist}
\end{figure}

\begin{figure}[!htb]
    \centering
    \includegraphics[scale=0.2]{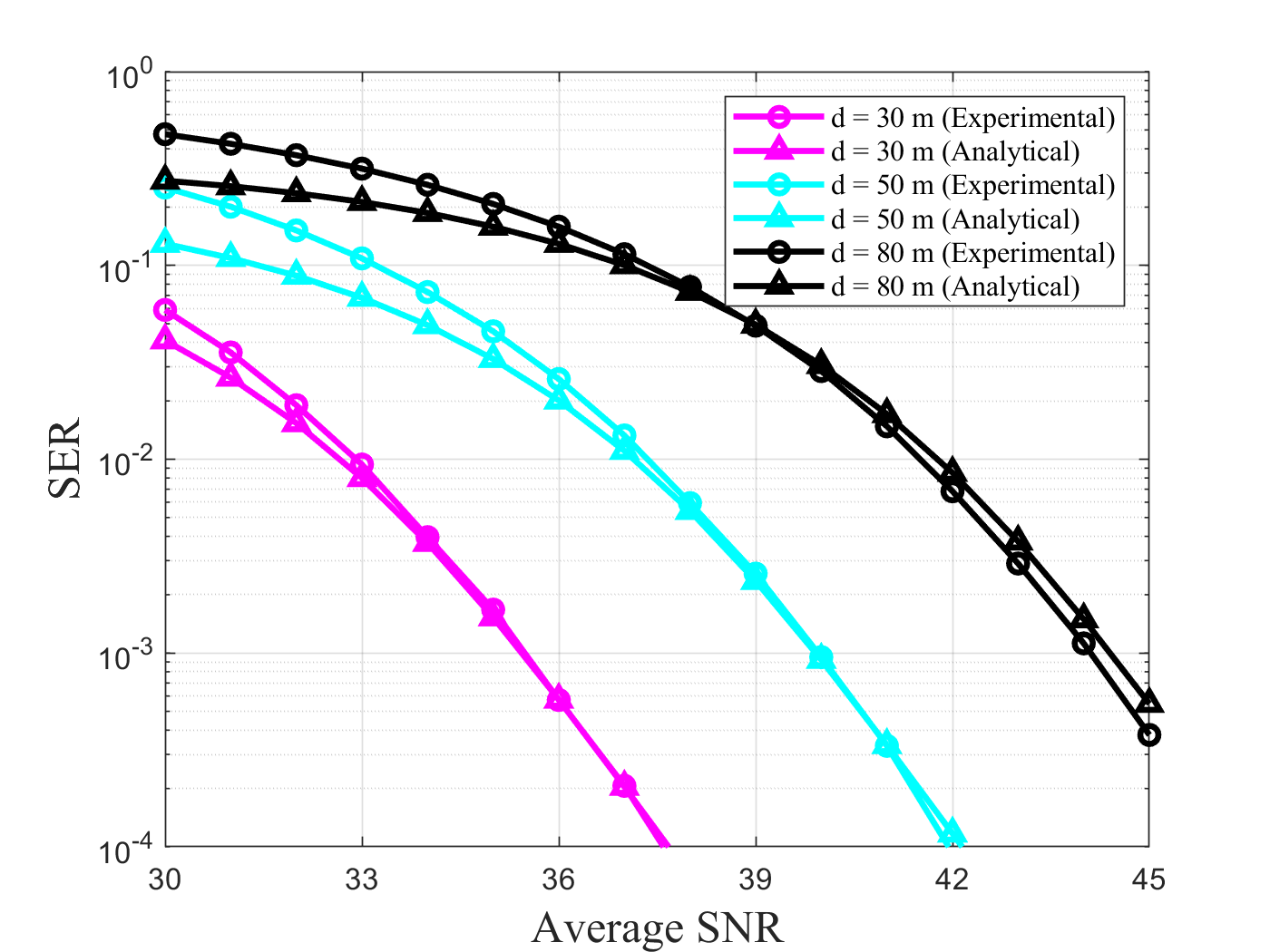}
    \caption{QPSK average SER vs. Average SNR assuming jitter variance of 0.01 $m^2$}
    \label{fig:dist_qpsk}
\end{figure}

Both deterministic and random fading components grow as functions of THz transmission distance. Therefore, it is convenient to compare theoretical and simulated system performance for variable range of transmission distances between 30 to 80 m. This is presented in Fig.  \ref{fig:dist} and Fig. \ref{fig:dist_qpsk} for BPSK and QPSK systems, respectively.

\begin{figure}[!htb]
    \centering
    \includegraphics[scale=0.2]{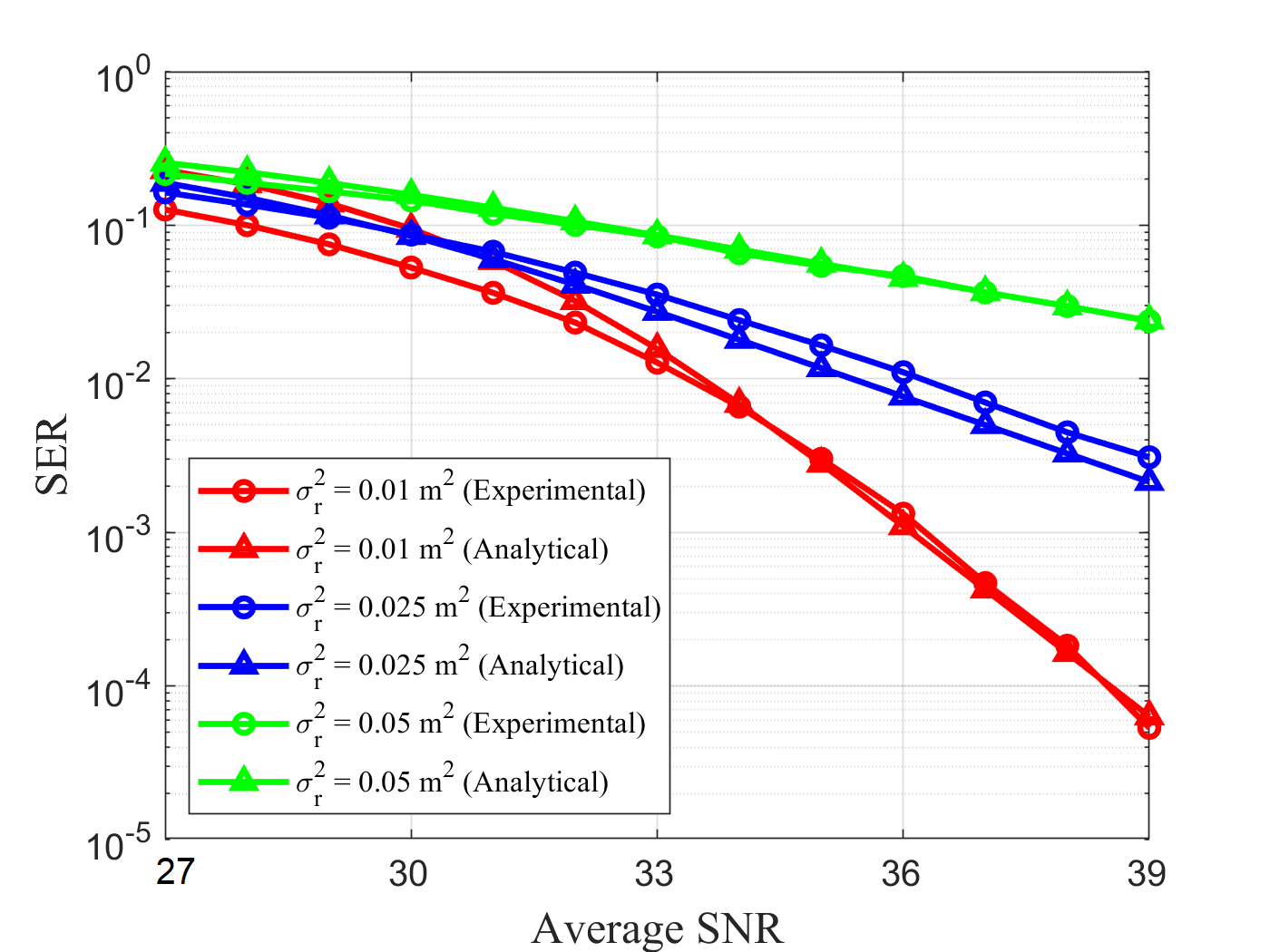}
    \caption{BPSK average SER vs. Average SNR assuming transmission distance of 50 m}
    \label{fig:vars}
\end{figure}

\begin{figure}[!htb]
    \centering
    \includegraphics[scale=0.2]{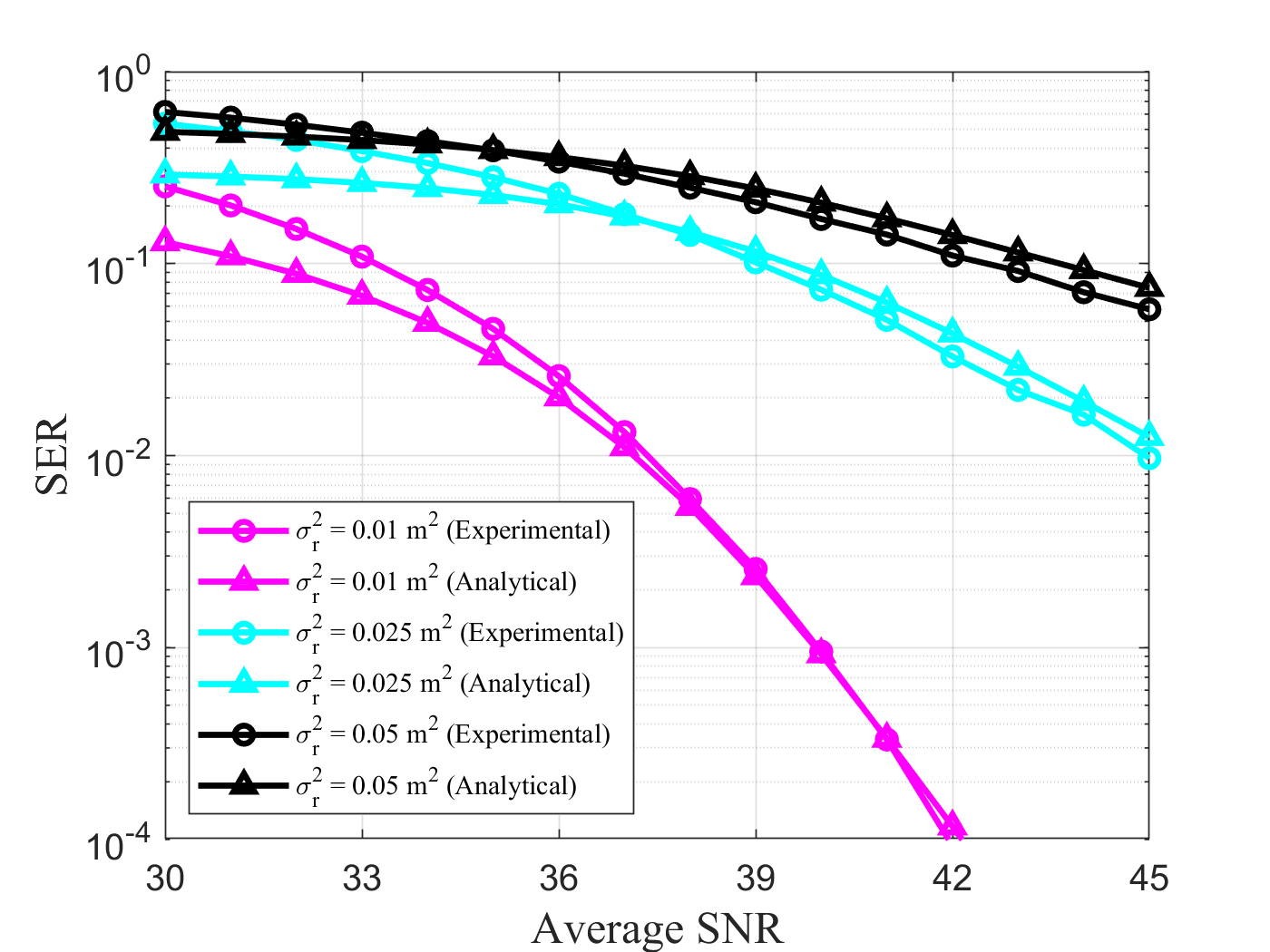}
    \caption{QPSK average SER vs. Average SNR assuming transmission distance of 50 m}
    \label{fig:vars_qpsk}
\end{figure}

Moreover, stochastic misalignment fading is quantified according to displacement jitter variance. Therefore, averaged  performance is also estimated assuming three jitter variance values of 0.01, 0.025 and 0.05 $m^2$, as shown in Fig. \ref{fig:vars} and Fig. \ref{fig:vars_qpsk} for BPSK and QPSK systems, respectively.  

 As expected, THz system performance severely degrades with both increasing transmission distance and displacement variance. Comparisons between theoretical error probabilities and Monte Carlo simulation results, for both BPSK and QBSK schemes, manifest good accuracy of the derived SER formulas for SNR levels above 35 dB. For lower SNR ranges, the derived SER expressions exhibit a maximum error of $\sim$2 dB SNR difference for distances above 30 m. For variable displacement variance, the error does not exceed 1 dB. It is important to note that the theoretical expressions (\ref{ser})  and (\ref{ser_qpsk}) are valid for a transmission distance range up to 100 m, maximum jitter variance of 0.05 $m^2$ and TX-RX beam radius ratio between 6 and 10. 
\section{Conclusion}\label{con}
The work in this paper considers statistical modeling and performance assessment of a LOS THz system applicable to short and medium range communications. The THz channel is characterized by studying the effects of propagation path loss, molecular absorption attenuation and transceivers antennas misalignment. Furthermore, approximated analytical formulas are derived to measure BPSK and QPSK system error probability as a function of average SNR and deterministic fading components. Theoretical performance results are verified against Monte Carlo simulations for different transmission distances and displacement variances. It was observed that the derived expressions are reasonably accurate for SNR levels above 35 dBs and yield a maximum of 2 dB mismatch error at SNR levels below 35 dB.

\bibliography{references}

\begin{thebibliography}{10}
\providecommand{\url}[1]{#1}
\csname url@samestyle\endcsname
\providecommand{\newblock}{\relax}
\providecommand{\bibinfo}[2]{#2}
\providecommand{\BIBentrySTDinterwordspacing}{\spaceskip=0pt\relax}
\providecommand{\BIBentryALTinterwordstretchfactor}{4}
\providecommand{\BIBentryALTinterwordspacing}{\spaceskip=\fontdimen2\font plus
\BIBentryALTinterwordstretchfactor\fontdimen3\font minus
  \fontdimen4\font\relax}
\providecommand{\BIBforeignlanguage}[2]{{%
\expandafter\ifx\csname l@#1\endcsname\relax
\typeout{** WARNING: IEEEtran.bst: No hyphenation pattern has been}%
\typeout{** loaded for the language `#1'. Using the pattern for}%
\typeout{** the default language instead.}%
\else
\language=\csname l@#1\endcsname
\fi
#2}}
\providecommand{\BIBdecl}{\relax}
\BIBdecl

\bibitem{chen2019survey}
Z.~Chen, X.~Ma, B.~Zhang, Y.~Zhang, Z.~Niu, N.~Kuang, W.~Chen, L.~Li, and
  S.~Li, ``A survey on terahertz communications,'' \emph{China Communications},
  vol.~16, no.~2, pp. 1--35, 2019.

\bibitem{liu2021thz}
S.~Liu, X.~Yu, R.~Guo, Y.~Tang, and Z.~Zhao, ``Thz channel modeling:
  Consolidating the road to thz communications,'' \emph{China Communications},
  vol.~18, no.~5, pp. 33--49, 2021.

\bibitem{kokkoniemi2016discussion}
J.~Kokkoniemi, J.~Lehtom{\"a}ki, and M.~Juntti, ``A discussion on molecular
  absorption noise in the terahertz band,'' \emph{Nano communication networks},
  vol.~8, pp. 35--45, 2016.

\bibitem{papasotiriou2020performance}
E.~N. Papasotiriou, A.-A.~A. Boulogeorgos, and A.~Alexiou, ``Performance
  analysis of thz wireless systems in the presence of antenna misalignment and
  phase noise,'' \emph{IEEE Communications Letters}, vol.~24, no.~6, pp.
  1211--1215, 2020.

\bibitem{oyeleke2020absorption}
O.~D. Oyeleke, S.~Thomas, O.~Idowu-Bismark, P.~Nzerem, and I.~Muhammad,
  ``Absorption, diffraction and free space path losses modeling for the
  terahertz band,'' \emph{Int. J. Eng. Manuf}, vol.~10, p.~54, 2020.

\bibitem{kokkoniemi2018simplified}
J.~Kokkoniemi, J.~Lehtom{\"a}ki, and M.~Juntti, ``Simplified molecular
  absorption loss model for 275--400 gigahertz frequency band,'' in \emph{12th
  European Conference on Antennas and Propagation (EuCAP 2018)}.\hskip 1em plus
  0.5em minus 0.4em\relax IET, 2018, pp. 1--5.

\bibitem{taherkhani2020performance}
M.~Taherkhani, Z.~G. Kashani, and R.~A. Sadeghzadeh, ``On the performance of
  thz wireless los links through random turbulence channels,'' \emph{Nano
  Communication Networks}, vol.~23, p. 100282, 2020.

\bibitem{dabiri2022pointing}
M.~T. Dabiri and M.~Hasna, ``Pointing error modeling of mmwave to thz
  high-directional antennas,'' \emph{arXiv preprint arXiv:2206.10756}, 2022.

\bibitem{farid2007outage}
A.~A. Farid and S.~Hranilovic, ``Outage capacity optimization for free-space
  optical links with pointing errors,'' \emph{Journal of Lightwave technology},
  vol.~25, no.~7, pp. 1702--1710, 2007.

\bibitem{boulogeorgos2019analytical}
A.-A.~A. Boulogeorgos, E.~N. Papasotiriou, and A.~Alexiou, ``Analytical
  performance assessment of thz wireless systems,'' \emph{IEEE Access}, vol.~7,
  pp. 11\,436--11\,453, 2019.

\bibitem{kokkoniemi2020impact}
J.~Kokkoniemi, A.-A.~A. Boulogeorgos, M.~Aminu, J.~Lehtom{\"a}ki, A.~Alexiou,
  and M.~Juntti, ``Impact of beam misalignment on thz wireless systems,''
  \emph{Nano Communication Networks}, vol.~24, p. 100302, 2020.

\bibitem{balanis2011modern}
C.~A. Balanis, \emph{Modern antenna handbook}.\hskip 1em plus 0.5em minus
  0.4em\relax John Wiley \& Sons, 2011.

\bibitem{alduchov1996improved}
O.~A. Alduchov and R.~E. Eskridge, ``Improved magnus form approximation of
  saturation vapor pressure,'' \emph{Journal of Applied Meteorology and
  Climatology}, vol.~35, no.~4, pp. 601--609, 1996.

\bibitem{miretti2022little}
L.~Miretti, T.~K{\"u}hne, A.~Schultze, W.~Keusgen, G.~Caire, M.~Peter,
  S.~Sta{\'n}czak, and T.~Eichler, ``Little or no equalization is needed in
  energy-efficient sub-thz mobile access,'' \emph{arXiv preprint
  arXiv:2210.05806}, 2022.

\bibitem{goldsmith2005wireless}
A.~Goldsmith, \emph{Wireless communications}.\hskip 1em plus 0.5em minus
  0.4em\relax Cambridge university press, 2005.

\bibitem{meghdadi2008ber}
V.~Meghdadi, ``Ber calculation,'' \emph{Wireless Communications}, 2008.

\bibitem{1210748}
M.~Chiani, D.~Dardari, and M.~Simon, ``New exponential bounds and
  approximations for the computation of error probability in fading channels,''
  \emph{IEEE Transactions on Wireless Communications}, vol.~2, no.~4, pp.
  840--845, 2003.

\end{thebibliography}
\bibliographystyle{IEEEtran}

\end{document}